\documentclass[letter]{aa} 

\usepackage{graphicx}
\usepackage{multicol}
\usepackage{longtable}
\usepackage{stfloats}

\usepackage{txfonts}

\usepackage[normalem]{ulem} 

\begin{document}

   \title{Fast-spinning massive black holes from slowly rotating low-metallicity stars: Implications for GW231123} 

\author{N.~H.~Ismail\inst{1,2}
        \and N.~Yusof\inst{1,2}\thanks{Corresponding author: norhaslizay@um.edu.my}
        \and R.~Hirschi\inst{3,4}
        \and A.~Griffiths\inst{3,5}
        \and M.~\'A.~Aloy\inst{5,6}
        \and S.~Ekstr\"om\inst{7}
        \and G.~Meynet\inst{7}
        }

\institute{
    Department of Physics, Faculty of Science, University of Malaya, 50603 Kuala Lumpur, Malaysia
    \and
    Center of Astronomy and Astrophysics, Faculty of Science, Universiti Malaya, 50603 Kuala Lumpur, Malaysia
    \and
    Astrophysics Research Centre, Lennard-Jones Laboratories, Keele University, Keele, ST5 5BG, UK
    \and
    Kavli IPMU (WPI), University of Tokyo, 5-1-5 Kashiwanoha, Kashiwa 277-8583, Japan
    \and
    Departament d’Astronomia i Astrofísica, Universitat de València, Av. Vicent Andrés Estellés 19, E-46100 Burjassot, València, Spain
    \and
    Observatori Astronòmic, Universitat de València, 46980 Paterna, Spain
    \and
    Geneva Observatory, University of Geneva, Chemin Pegasi 51, CH-1290 Sauverny, Switzerland
}

   \date{Received : xxx}

\abstract
%
{The origin of massive black holes in the early Universe remains uncertain and is still unexplored. Population III (Pop III; zero-metallicity) stars are among the first stellar sources capable of producing such remnants, but their evolution is very sensitive to rotation.}
{We explore how slow initial rotation affects the evolution and black hole formation of very massive Pop III stars and assess their potential to become massive fast-spinning black holes consistent with gravitational-wave events such as GW231123.}
%
{We computed a grid of non-rotating and slowly rotating Pop III stellar models with initial masses of 80, 85, and $90~M_\odot$ using the \texttt{GENEC} stellar evolution code. Our models included rotation-induced mixing and angular momentum transport by magnetic torques (Taylor-Spruit dynamo and the magneto-rotational instability). We analysed the CO core masses and their volume-averaged adiabatic index $\langle \Gamma_1 \rangle$ to assess stability against 
electron-positron pair creation.
From the angular momentum {}{profiles of the stellar models by the end of He-burning, we} estimated the resulting black hole masses and dimensionless spins under the assumption of a direct collapse.}
{Our non-rotating and slowly rotating 80 and $85~M_\odot$ models 
developed carbon-oxygen core masses between $31\text{and }36\,M_{\odot}$ and had an adiabatic index $\langle\Gamma_{1}\rangle$, which remained above $4/3$  
(i.\,e. they avoided pair instability).

Our models thus predict that Pop III stars can
keep most of their mass and collapse directly to form black holes of $80\text{--}85\,M_{\odot}$ with dimensionless spins up to $a_{\mathrm{BH}} \lesssim 0.7$.}
{Initially slowly rotating massive Pop III stars can form very massive rapidly spinning black holes just below the pair-instability regime. This supports an interpretation of the lower boundary of the pair-instability supernova  mass gap as a smooth, structure-dependent transition and identifies single-star Pop III evolution as 
{}{a possible component channel for} the population of massive 
fast-spinning
black holes observed by gravitational-wave detectors{, subject to the uncertain efficiency of internal angular momentum transport and mass-loss prescriptions}. }


\authorrunning{N.H Ismail et al.}
\titlerunning{Fast-spinning BH from slow Pop III rotators}
\maketitle
\nolinenumbers

\def\showmaasection{} 

\ifdefined\showmaasection
\section{Introduction}

The gravitational-wave event GW231123 has been interpreted as the merger of two black holes (BHs) with source-frame masses of $137^{+23}_{-18}\,M_\odot$ and $101^{+22}_{-50}\,M_\odot$, and with high inferred spins, $\chi_1=0.90^{+0.10}_{-0.19}$ and $\chi_2=0.80^{+0.20}_{-0.52}$ \citep{Abac2025}. These properties make GW231123 particularly challenging to interpret within standard stellar evolution channels. In conventional models, BH formation is expected to be strongly suppressed in the pair-instability mass gap (PIMG), which is commonly placed at roughly $\sim 55$--$130\,M_\odot$ \citep{Heger2003,Giacobbo2018,Woosley2019,Woosley_2021}. The physical origin of this gap lies in the development of sufficiently massive carbon--oxygen (CO) cores. Within such cores, pair creation softens the equation of state, triggering contraction and explosive oxygen burning, which can either eject part of the stellar envelope in pulsational pair-instability (PPI) episodes or disrupt the star completely in a pair-instability supernova (PISN), leaving no compact remnant \citep{Heger2003,Farmer_2019,Woosley_2021}.

GW231123 is especially intriguing in this context. Its secondary component lies well within the canonical PIMG, while the primary lies within or above its upper edge, depending on the true source parameters. In addition to their high masses, both components appear to be spinning rapidly \citep{Abac2025}. It is therefore a non-trivial challenge for single-star evolutionary models to simultaneously explain such high masses and high spins.

A number of scenarios have been proposed to explain BHs in the mass gap, including growth of lower-mass BHs through accretion or repeated mergers \citep{Roupas2025,Steigman2025,Kiroglu2025,Liu2025,Gottlieb2025}, premature collapse before pair instability fully develops \citep{Baumgarte2025}, and a primordial origin \citep{Nojiri2025,Yuan2025}. These possibilities suggest that the PIMG is not an immutable prediction, but the outcome of competing effects involving core growth, mass loss, rotational mixing, angular momentum transport, and the thermodynamic response of the stellar interior \citep{Farmer_2019,Renzo_2020,Woosley_2021}.

The difficulty of forming BHs in the mass gap through classical stellar evolution stems from the relation between the final stellar mass and the CO-core mass, since the latter determines the onset of pair instability (PI). This relation is sensitive to mass loss, chemical mixing, rotation, convection, and nuclear reaction rates.

Recent work suggested that the lower edge of the PPI regime is set primarily by a critical CO-core mass, with residual dependence on rotation and mass loss \citep{winch_2025}. Using the same stellar evolution code as in the present work, \citet{Farrell2021} found that Population~III stars may form BHs of up to about $85\,M_\odot$; this was later supported by \citet{Hirschi_2025} for very low-metallicity models ($Z=10^{-5}$).

Population~III stars are particularly relevant here as their primordial composition implies extremely weak radiatively driven winds, allowing them to retain most of their mass and potentially much of their angular momentum throughout their evolution. They are therefore promising progenitors of very massive possibly rapidly spinning BHs, and they are natural candidates for probing the lower edge of the PIMG \citep{Heger2003,Farrell2021,Hirschi_2025}.

This motivated us to re-examine the lower edge of the PIMG at zero metallicity, with special emphasis on the role of slow rotation and angular momentum retention. Instead of determining the highest BH mass attainable below the onset of PI, we determine the maximum mass of a fast-spinning BH that can be produced by a single very low-metallicity star. To address this issue, we computed new stellar-evolution models of %
massive Pop~III stars with initial masses of 80, 85, and $90\,M_\odot$ using the \texttt{GENEC} code. We followed the evolution of non-rotating and slowly rotating models and analysed their mass retention, core growth, angular momentum budget, and stability against PI using their CO-core masses and volume-averaged adiabatic index. Our aim was to determine whether stars in this mass range can avoid pair-instability disruption and collapse into massive BHs, and whether the resulting remnants can retain sufficiently high spins to be relevant for systems such as GW231123.

The paper is organised as follows. In Sect.~2 we describe the stellar models and summarise their main evolutionary properties. In Sect.~3 we discuss their predicted fate and the implications for BH masses and spins near the pair-instability boundary. The summary of the evolutionary properties is presented in Appendix~\ref{sec:evolutionary table}. An approximate treatment of collapse, fallback, and BH growth is presented in Appendix~\ref{sec:fate_prediction}.

\section{Stellar models}

\vspace{1em}

We computed the evolution of 
massive stars for three initial masses (80, 85, and $90~M_\odot$) at zero metallicity ($Z = 0$) from the zero-age main sequence (ZAMS) to at least the end of core-helium burning and generally until oxygen burning using the stellar evolution code GENEC \citep{RN915}.  
The models were evolved with varying degrees of initial rotation, from non-rotating to a critical velocity of 20\% at most. The specific initial values and nomenclature for all models we computed can be found in Table~\ref{tab:Z0_models_BH}. The choice of masses and rotation was motivated by the predicted high mass and high spin of the BH remnant inferred from the event GW231123 (and the expected lower boundary of the PIMG). We used zero-metallicity models to minimise mass loss from stellar winds  and to explore first stellar generation BHs. 

In metal-free stars, line-driven winds are extremely weak, so any significant 
loss of mass, and thus, of total angular momentum, can only occur through mechanical mass shedding triggered by proximity to the critical surface rotation. The initial rotational velocities we chose are relatively low to avoid such mechanical mass loss. This is in contrast with previous studies with GENEC (cf. \cite{Murphy_2021} uses $\omega=v_{\rm ini}/v_{\rm crit}=0.4$, where $v_{\rm crit}$ is the critical velocity). The suppression of mechanical and wind-driven mass loss left our models with a high final mass and total angular momentum, making them good candidates to produce heavy  BHs {that can be spun up by the late accretion of angular-momentum-rich outer layers}.

We used the version of GENEC described in \cite{griffiths2025evolving}, including the same physics as in \cite{ekstrom2012grids} and the recent grid of \cite{Sibony_2024}, except for three key aspects. These consist of updates to the equation of state (EoS), opacity calculations, and nuclear reaction network (we employed \texttt{GeValNet25} ; for further details on the differences, see Sections 2, 3, and 4 of \cite{griffiths2025evolving}). These updates, particularly that of the EoS, are critical to correctly determine the evolution of the core of the massive stars under thermodynamic conditions where the pair-instability becomes important. 

Our models that evolved with rotation also included {angular-momentum} 
transport  induced by {all standard shear instabilities, meridional 
circulation, convection \citep[see,][for a complete list of 
mechanisms included in GENEC]{RN915}, and} local magnetic torques. 
{The magnetic mechanisms of angular momentum transport we used}
{are} the Tayler-Spruit (TS) dynamo, as described in 
\cite{Eggenberger_2022}, {and} the magneto-rotational instability 
(MRI), as described in \cite{Griffiths_2022}. These two 
instabilities {can} transport angular momentum {outwards within the 
stellar interior} during the main sequence and core-helium burning.
{The efficiency of these prescriptions remains uncertain, and there 
is no observational evidence that these magnetic instabilities 
operate in Pop III stars. Therefore, the models should be 
interpreted as exploring one possible (perhaps extreme) scenario of angular- momentum transport rather than establishing the probability 
of this channel.}
 {Angular momentum transport shapes} the final distribution of 
 angular momentum within the stellar interior {since our baseline 
 models effectively lost no mass (see below), and hence, conserved the 
 total angular momentum}.

In Table~\ref{tab:Z0_models_BH} we provide the values for the 
initial, $M_{\rm ini}$, and final mass, $M_{\rm fin}$ (i.e. at the 
end of core-helium burning) and the variation with respect to the 
initial mass, $\Delta M$. Only the models with 
$M_{\rm ini}>85M_{\odot}$ and $\omega\ge 0.20$ underwent appreciable 
mass loss due to high surface rotation during the stellar expansion 
after the main sequence. Mechanical mass loss is therefore avoided 
provided the initial rotation stays below $\approx20\%$ of the 
critical velocity. We also report the total angular momentum at the ZAMS 
and at the end of core-helium burning, which again only decreased 
appreciably for models with  $\omega= 0.20$,
as all other models retained nearly all of their mass, and 
consequently, their angular momentum. Finally, we provide the values 
of the helium core mass, $M_{\rm \alpha}$, defined as the location 
where the hydrogen mass fraction falls below 0.01 and of the CO core 
mass, $M_{\rm CO}$, defined as the location where the helium mass 
fraction falls below 0.01. These two masses are evaluated at the 
end of core-helium burning. The carbon-oxygen core mass might grow 
slightly during later evolutionary phases, but the values we report 
approximately correspond to the final core masses at collapse. 
$M_{\rm \alpha}$ and $M_{\rm CO}$ are commonly used to predict 
whether a star will undergo a PISN. CO core masses below 
$60\,M_{\odot}$ \citep{Hirschi_2017,Woosley_2017} are not expected 
to produce a full PISN. For cores in the range $\sim 40$–
$60\,M_{\odot}$, violent pair-instability pulsations can occur, 
leading to eruptive mass loss, but the pulsations are not strong 
enough to completely disrupt the core. In our models, $M_{\rm 
\alpha}$ and $M_{\rm CO}$ were below the lower limit of 
$40\,M_{\odot}$, except for $M_{\rm \alpha}$ in  models with $M_{\rm 
ini}=90\,M_\odot$,  
for which they were marginally above $40\,M_{\odot}$.  We further  
assessed whether the models presented here were globally dynamically 
stable to pulsations. {To do this, most of our models were 
evolved beyond core-helium burning (see 
Table\,\ref{tab:Z0_models_BH}).} The models were globally stable when the 
adiabatic index $\Gamma_1$ satisfied the following inequality:
\begin{equation}
    \langle \Gamma_1 \rangle =\frac{ \int \Gamma_1 \frac{P}{\rho} \rm dm}{\int  \frac{P}{\rho} \rm dm} > 4/3,
\label{eq:G1}
\end{equation}
as already used in \cite{Renzo_2020} and based on the original derivation of this stability criterion by \citet{Ledoux_1945} and \citet{Stothers_1999}. The models that remained globally stable (became globally unstable) until oxygen burning are labelled with an N (Y) in Table\,\ref{tab:Z0_models_BH}. 
 In Appendix~\ref{sec:fate_prediction} we describe the method for predicting the final remnant parameters of our models.

Figure~\ref{fig:Kipp_P080z00S005_ageadv_Mr} shows the Kippenhahn 
diagram of the $80\,M_\odot$, with an initial velocity ratio 
$\omega=0.05$, illustrating the internal structure evolution from 
the main sequence to core oxygen burning. The total stellar mass 
remains nearly constant throughout the evolution, as previously 
discussed. Convective regions are confined to the core during H and 
He core burning, while the envelope stays radiative. This internal structure 
is consistent with the moderate helium and carbon-oxygen core masses 
reported in Table~\ref{tab:Z0_models_BH} and characterises an 
evolution that avoids strong structural instabilities prior to core 
collapse. By retaining most of its mass and angular momentum, 
especially in the outer layers   (50\% of the angular momentum is 
contained in the outer 4--5\,$M_\odot$; see the vertical line in 
Fig.\ref{fig:Max_final_a}), the star can evolve towards direct collapse and 
subsequently spin the BH up through fallback accretion.

\begin{figure}
    \centering
    \includegraphics[width=\linewidth]{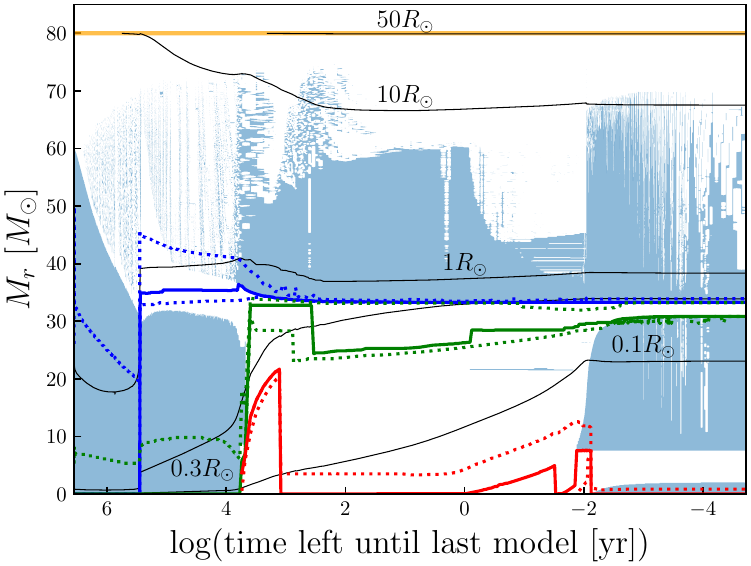}
    \caption{Kippenhahn diagram of the $80\,M_\odot$ model with an initial rotation set to $5\%$ of the critical velocity.  
    The black lines denote radius contours at $0.1$, $0.3$, $1$, $10$, and $50\,R_\odot$, and the orange line marks the total mass of the star, which is constant for this model. The light blue shaded regions correspond to convective zones. 
    The coloured lines show the locations of the peak (solid lines) and of $~10\%$ of the peak (dotted lines) in the nuclear energy generation rate for hydrogen burning (blue), helium burning (green), and advanced burning (carbon, neon, or oxygen; red).}
    \label{fig:Kipp_P080z00S005_ageadv_Mr}
\end{figure}

\section{Discussion and conclusions}

The PPI boundary sets the maximum stellar mass that can collapse to a BH without being disrupted (below the PIMG). This limit is not determined by a unique initial mass, but by the onset of dynamical instability when electron--positron pair production softens the EoS in sufficiently massive CO cores. Early stellar evolution and hydrodynamical studies placed the onset of pulsational pair instability at CO-core masses of roughly $35$--$40\,M_\odot$ \citep{Heger_2002,Farmer_2019,Woosley2019}, while more recent systematic grids suggest a nearly constant critical value of $M_{\rm CO,crit}\approx35$--$36\,M_\odot$ \citep{winch_2025}, with remaining uncertainties linked to nuclear reaction rates and convection modelling \citep{Renzo_2020,Farag_2022,Woosley_2021}. Below this threshold, stars are expected to collapse directly to BHs; above it, they can enter {}{a transitional regime in which weak or moderate pulsational pair-instability (PPI) episodes can occur before final collapse, whereas complete PISN disruption requires substantially larger cores ($60\,M_\odot \lesssim M_{\rm CO}\lesssim 130\,M_\odot$)}.

Within this framework, our Pop~III models approached the PPI boundary from below. The $80$ and $85\,M_\odot$ sequences developed CO-core masses between $\approx31$ and $35.9\,M_\odot$, that is, at or below the nominal instability threshold. 

For these $80$ and $85\,M_\odot$ models, the volume-averaged adiabatic index satisfies $\langle\Gamma_1\rangle > 4/3$, and they thus show no evidence of global dynamical instability. These models are therefore expected to collapse directly, producing BHs with masses up to $\sim80$--$85\,M_\odot$.

By contrast, the $90\,M_\odot$ models reach CO-core masses of 
$\sim36.5$--$38\,M_\odot$, placing them in the transitional regime 
commonly associated with the onset of pulsational pair-instability 
mass ejection episodes or supernovae \citep[sometimes referred to as 
PPISNe, which are less violent than the the full PISNe;][]
{Farmer_2019,winch_2025}. Although global instability has not been 
explicitly confirmed for all our models, their core masses fall 
within the range where PPISN is expected, depending on the details 
of the late evolutionary stages. 
We stress that placing these models near the lower PI boundary does 
not imply complete disruption; rather, it indicates that pulsational 
pair-instability effects can begin to affect the final mass and 
structure. This reinforces the view that the PPI boundary is better 
interpreted as a structural transition region and not as a sharp 
cut-off.

The most distinctive result of this work concerns angular momentum 
retention and its effect on the final BH spin. 
Because metal-free models have negligible ordinary wind-driven mass 
loss, angular momentum transported toward the surface cannot be 
efficiently removed and instead accumulates in the outer envelope. 
These layers remain bound provided the star does not rotate rapidly 
enough, $\omega\lesssim0.15$, for centrifugal ejection to occur. As 
a consequence, the BH that forms is initially close to non-rotating 
because the inner stellar regions have been efficiently drained of 
angular momentum{}{; magnetic torques are crucial for this redistribution}. {The high final BH spin is therefore not inherited 
from an initially rapidly rotating core, but is produced later by 
the accretion of outer layers where most of the stellar angular 
momentum is stored  (Fig.\,A1 illustrates this point for the model 
with $M_{\rm ini}=85\,M_\odot$ and $\omega=0.15$; a qualitatively 
similar behaviour is found in the other models)}. This strongly 
limits the feedback exerted by the newly formed BH until the 
outermost layers, in which most of the angular momentum is stored,
begin to accrete. The BH can therefore grow substantially in mass 
before its spin increases significantly. This behaviour contrasts 
with that of rapidly rotating progenitors, for which collapse can 
produce fast-spinning BHs whose feedback can strongly reduce the 
final remnant mass or even disrupt the star \citep[e.g.][]
{Shibata_2026ApJ...996...57}.

Using the simplified one-dimensional collapse and feedback model 
(Appendix~\ref{sec:fate_prediction}), we estimated the fraction of mass 
and angular momentum retained during disk formation and fallback. 
Even under extreme assumptions, disk-driven outflows remove %
$10\%$ of the stellar mass at most. The final BH mass therefore retains most of the progenitor mass.

The final spin reflects the interplay between {}{three factors:  (i) outward angular momentum redistribution during stellar evolution, (ii) weak angular momentum loss at zero metallicity, and (iii) late accretion of material with high angular momentum}. In our model grid, the dimensionless spin parameter $a_{\rm BH}$ increased with the initial rotation parameter $\omega$ to a maximum and then decreased, reaching peak values of $a_{\rm BH}\simeq0.7$ for $M_{\rm ini}\simeq85\,M_\odot$. This non-monotonic behaviour shows that rapid BH rotation does not require extreme initial stellar rotation. Rather, even moderately rotating Pop~III stars can produce high-mass and rapidly spinning remnants when winds are negligible and the progenitor remains below the PPI boundary. 

 While the inferred redshift of GW231123 is moderate ($z \sim 0.4$ see \cite{Abac2025}) and the metallicity of its BHs might be higher than zero, the findings of this study are  consistent with stellar grids predicting BH masses up to $\sim80$--$90\,M_\odot$ just below the PPI regime at very low metallicity \citep{Hirschi_2025}. These results thus provide a useful point of comparison with GW231123 (assuming its BHs formed from very low-metallicity stars), whose components have inferred masses and spins of $M_1 = 137^{+23}_{-18}\,M_\odot$, $a_1 = 0.90^{+0.10}_{-0.19}$, and $M_2 = 101^{+22}_{-50}\,M_\odot$, $a_2 = 0.80^{+0.20}_{-0.52}$ \citep{Abac2025}. 

We stress, however, that our calculations are single-star models and do not address the binary assembly of GW231123. In particular, we did not model tidal interactions, mass transfer, envelope stripping, or the orbital separations required to preserve the hydrogen envelope while still producing a binary black hole that merges within a Hubble time.

Within the mass range we explored, only the properties of the secondary BH can be approached{}{, and only from the lower-mass side}. Our models produce BHs with masses up to $\sim80$--$90\,M_\odot$ and spin them up to $a_{\rm BH}\simeq0.7$, which is slightly below but still broadly consistent within present uncertainties with the inferred spin of the secondary. The substantially more massive primary component lies beyond the progenitor masses considered here and likely requires either more massive progenitors above the PIMG or a different formation channel. {The connection between such single-star remnants and a merging binary system, whether through isolated binary evolution or dynamical pairing, must therefore be investigated in future work.} More generally, gravitational-wave population studies suggest a downturn in the BH mass distribution between $\sim45$ and $60\,M_\odot$, commonly associated with the PI mass gap \citep{Fishbach_2017,Farr_2019}, although more recent analyses do not require a sharp cut-off below $\sim40$--$50\,M_\odot$ \citep{Ray_2026}. This is consistent with interpreting the PPI boundary as a transition region and not as an abrupt limit. \\

\indent In summary, the PPI boundary acts as a physical ceiling for single-star BH formation. Below this ceiling, however, slowly rotating metal-free stars can naturally produce high-mass and rapidly spinning BHs, provided that mass loss remains negligible and angular momentum is efficiently retained. 
The combination of weak mass loss, the internal angular momentum redistribution, and the retention of most of the stellar mass during collapse {}{suggests that within the prescriptions for angular momentum transport we adopted,} high-mass fast-spinning BHs in the early Universe did not require extreme initial rotation, but {}{might} arise  from massive Pop~III progenitors evolving just below the PPI boundary.

\begin{acknowledgements}
 
  NY and NHI acknowledges the Fundamental Research Grant Scheme 
  grant number  FRGS/1/2021/STG07/UM/02/4 under Ministry of Higher 
  Education, Malaysia. MAA and AG have been supported  from the 
  grant PID2021-127495NB-I00 and PID2025-171322NB-C22, funded by 
  MCIN/AEI/10.13039/501100011033 and by the European Union under 
  ``NextGenerationEU'', the Astrophysics and High Energy Physics 
  program of the Generalitat Valenciana ASFAE/2022/026 funded by 
  MCIN and the European Union ``NextGenerationEU'' (PRTR-C17.I1), 
  and the Prometeo excellence program grant CIPROM/2022/13 funded by 
  the Generalitat Valenciana. RH acknowledges support from STFC, the 
  World Premier International Research Centre Initiative (WPI 
  Initiative), MEXT, Japan, the IReNA AccelNet Network of Networks 
  (NSF, Grant No. OISE-1927130), CeNAM (grant DE-SC0026204) and {the 
  Wolfson Foundation that part-funded the greenHPC facility at 
  Keele}. 
\end{acknowledgements}

%
%
\bibliographystyle{aa} 
\bibliography{vms_z0} 
%
%

\begin{appendix}

\section{Stellar grid summary}
\label{sec:evolutionary table}
This appendix provides comprehensive summary of the evolutionary 
outcomes and predicted black hole (BH) remnants for both non-
rotating and rotating stellar models at zero metallicity $Z=0$. 
Table \ref{tab:Z0_models_BH} tracks key parameters  of  Population III stars, from the zero-age main sequence (ZAMS) through the end of core-helium burning.

\setlength\tabcolsep{3.5pt}
\renewcommand{\arraystretch}{1.3}
\begin{table*}[!ht]
\centering
\caption{
Evolutionary outcomes and BH remnant predictions for non-rotating and rotating $Z=0$ stellar models. }
\label{tab:Z0_models_BH}
\begin{tabular}{lccccccc|cc|cccccc}
\hline\hline
$M_{\mathrm{ini}}$ &
$\omega$ &
$J_{\mathrm{ini}}$ &
$M_{\mathrm{fin}}$ &
$J_{\mathrm{fin}}$ &
$M_{\alpha}$ &
$M_{\mathrm{CO}}$ &
$\Delta M$ &
$\mathrm{final}$ &
$\mathrm{PPI}$  &
$M_{\mathrm{BH}}^{(1)}$ &
$M_{\mathrm{BH}}^{(2)}$ &
$M_{\mathrm{e}}^{(1)}$ &
$M_{\mathrm{e}}^{(2)}$ &
$a_{\mathrm{BH}}^{(1)}$ &
$a_{\mathrm{BH}}^{(2)}$ \\
 & [\%] & [g\,cm$^2$\,s$^{-1}$] & [$M_\odot$] & [g\,cm$^2$\,s$^{-1}$] & [$M_\odot$] & [$M_\odot$] & [$10^{-4}\,M_\odot$] & phase & Y/N & [$M_\odot$] & [$M_\odot$] & [$M_\odot$] & [$M_\odot$] &  &  \\
\hline
80   & 0  & —                & 79.9999 & —                & 35.70 & 31.78 & 1.10 & O     & N & 80 & — & 0.0 & — & 0.00 & — \\
80   & 5  & 3.54$\times$10$^{52}$ & 79.9999 & 3.53$\times$10$^{52}$ & 34.76 & 31.30 & 1.07 & O     & N & 79.7 & 78.6 & 0.3 & 1.4 & 0.41 & 0.38 \\
80   & 10 & 7.06$\times$10$^{52}$ & 79.9999 & 7.06$\times$10$^{52}$ & 34.87 & 30.88 & 1.21 & He    & `N' & 79.1 & 76.4 & 0.9 & 3.6 & 0.67 & 0.61 \\
80   & 15 & 1.05$\times$10$^{53}$ & 79.9999 & 1.05$\times$10$^{53}$ & 35.97 & 31.99 & 1.31 & He    & `N' & 78.9 & 72.0 & 1.1 & 8.0 & 0.64 & 0.48 \\
\hline
85   & 0  & —                & 84.9999 & —                & 38.01 & 34.89 & 1.12  & O    & N & 85 & — & 0.0 & — & 0.00 & — \\
85   & 5  & 3.93$\times$10$^{52}$ & 84.9999 & 3.93$\times$10$^{52}$ & 38.22 & 35.49 & 1.18 & He   & `N' & 84.9 & 84.1 & 0.1 & 0.9 & 0.38 & 0.36 \\
85   & 10 & 7.86$\times$10$^{52}$ & 84.9999 & 7.85$\times$10$^{52}$ & 38.43 & 34.71 & 1.29 & O    & N & 84.2 & 80.5 & 0.8 & 4.5 & 0.54 & 0.44 \\
85   & 15 & 1.17$\times$10$^{53}$ & 84.9999 & 1.17$\times$10$^{53}$ & 38.15 & 33.84 & 1.47 & He   & `N' & 83.7  & 80.9 & 1.3 & 4.1 & 0.70 & 0.65 \\
85   & 20 & 1.56$\times$10$^{53}$ & 84.7772 & 1.14$\times$10$^{53}$ & 39.89 & 35.97 & 2.23 $\times10^3$ & He & `N' & 83.7 & 77.6 & 1.1 & 7.2 & 0.61 & 0.46 \\
\hline
90   & 0  & —                & 89.9999 & —                & 40.77 & 37.09 & 1.20 & O      & Y   & (90) & — & (0.0) & — & (0.00) & — \\
90   & 5  & 4.36$\times$10$^{52}$ & 89.9999 & 4.36$\times$10$^{52}$ & 40.39 & 36.54 & 1.30 & He-C  & `Y' & (89.6) &  (89.6) & (0.5) & (0.5) & (0.36) & (0.36) \\
90   & 10 & 8.69$\times$10$^{52}$ & 89.9999 & 8.69$\times$10$^{52}$ & 40.92 & 36.65 & 1.38 & Ne    & `Y' & (89.0) & (87.3) & (1.0) & (2.7) & (0.66) & (0.63) \\
90   & 15 & 1.30$\times$10$^{53}$ & 89.9983 & 1.29$\times$10$^{53}$ & 41.67 & 37.67 & 1.65$\times10^1$ & He & `Y' & (88.9) & (80.5) & (1.1) & (9.5) & (0.60)  & (0.41) \\
90   & 20 & 1.72$\times$10$^{53}$ & 89.7474 & 1.24$\times$10$^{53}$ & 42.46 & 37.87 & 2.53$\times10^3$ & H-He & `Y' & (88.6) & (81.1)  & (1.1)  & (8.7) & (0.61) & (0.45) \\
\hline
\end{tabular}
\tablefoot{The table tracks key parameters from the zero-age main sequence (initial mass, rotation $\omega\equiv v_{\mathrm{ini}}/v_{\mathrm{crit}}$ and angular momentum, $J_{\mathrm{ini}}$) to the end of core helium-burning (final mass $M_{\mathrm{fin}}$, final angular momentum $J_{\mathrm{fin}}$, He core mass $M_{\alpha}$, CO core mass $M_{\mathrm{CO}}$, and total mass loss $\Delta M$). 
{The column labelled final phase gives the final burning phase reached by each model, and indicates how far the evolution has been followed when diagnosing whether the model becomes globally PPI unstable (O, Ne, C, and He denote core oxygen, neon, carbon, and helium burning).}
The column labelled PPI indicates whether the model is unstable according to the global pair-instability criterion (Eq.\,\ref{eq:G1}).
A Y indicates models that become globally PPI unstable.
An N denotes models that are stable against the PPI. 
A Y {or N} with quotation marks (`Y' {or `N'}) denotes models that have not been evolved sufficiently far to confirm global dynamical instability explicitly, but whose CO-core masses place them within the expected transitional PPI regime.
The last six columns provide the masses and dimensionless spins of the BHs formed from these models, along with the ejected mass, $M_{\rm e}$ according to the two extreme feedback cases described in App.\,\ref{sec:fate_prediction}. Numbers enclosed in parentheses denote values computed for marginal cases which are likely to enter the PPI regime.}
\end{table*}

\section{BH growth and feedback model}
\label{sec:fate_prediction}

In stellar models whose helium core mass hints at the formation of a 
BH, we follow the BH growth and feedback onto the rest of the star 
adopting the semi-analytic framework of 
\citet{Batta_2019arXiv190404835}, with minor modifications. 
Accretion proceeds shell-by-shell. Material with equatorial specific 
angular momentum, $j_{\rm eq}$ above the last stable orbit (LSO) 
threshold forms a disk and can power winds; otherwise, it accretes 
quasi-radially. Feedback is compared to the binding energy of the 
exterior layers to determine whether the envelope is unbound and BH 
growth terminates.

From stellar profiles ($r,\rho,P,\Gamma,\Omega$) we compute shell 
masses $\Delta m_i=M_{r,i}-M_{r,i-1}$ and $j_{{\rm eq},i}=\Omega_i 
r_i^2$. For a BH of mass $M_{\rm BH}$ and spin $a_{\rm BH}$, shells 
with $j_{{\rm eq},i}\le j_{\rm LSO}$ plunge into $r_{\rm LSO}$, while 
those with $j_{{\rm eq},i}>j_{\rm LSO}$ circularize at $r_{\rm 
circ}=j_{{\rm eq},i}^2/(GM_{\rm BH})$ and form a disk spanning 
$\sin\theta_{d,i}=\sqrt{j_{\rm LSO}/j_{{\rm eq},i}}$. Following 
\citet{Bardeen_1970Natur.226...64} and 
\citet{Thorne_1974ApJ...191..507}, disk-accreted mass $\Delta 
m_{{\rm disk},i}$ adds energy, $e_{\rm LSO}(a_{\rm BH})$ and specific 
angular momentum, $j_{\rm LSO}$, whereas the quasi-radial fraction 
$\Delta m_{{\rm radial},i}$ adds mass and specific angular momentum 
$A_p j_{{\rm eq},i}$, with $A_p=(2/3)(1-\cos^3\theta_{d,i})/(1-
\cos\theta_{d,i})$. We assume BH formation at $M_{{\rm 
BH},0}=3\,M_\odot$ with spin $a_{{\rm BH},0}=GJ_0/(c^2M_{{\rm 
BH},0}^2)$, where $J_0$ is the total angular momentum enclosed in 
$M_{\rm BH,0}$. {In our models this seed spin is small; the 
substantial spin-up occurs later, as higher-angular-momentum outer 
shells are accreted.}

Disk winds inject energy at a rate $\dot E_{\rm wind}=\eta_w(a_{\rm BH})\,\varepsilon\,\dot M c^2$, where $\eta_w\simeq 0.05+0.35a_{\rm BH}^2$ \citep{McKinney_2012MNRAS.423.3083} and $\varepsilon=(v_{\rm wind}/c)^2$ with $v_{\rm wind}=\mathcal{M}_{\rm wind} c_s$, $\mathcal{M}_{\rm wind}$ the wind Mach number and $c_s$ the sound speed in the vicinity of the BH. Only disk-accreted mass contributes to feedback, so shell $i$ injects
$E_{\rm fb}^{(i)}=\eta_w(a_{\rm BH})\,\varepsilon\,\Delta m_{{\rm disk},i}\,c^2$.

{We use a dynamic energy barrier, in which shell $i$ is assumed to have fallen from its original radius $r_i$ to an arrival radius $\tilde r_i$ before being stalled by feedback. The feedback must therefore both cancel the infall kinetic energy of shell $i$ and lift it out of the potential from $\tilde r_i$, in addition to unbinding the exterior layers. The corresponding barrier is}
\begin{align}
E_{\rm ub}^{(i)}=&\frac{1}{2}\Delta m_i v_i^2({}{\tilde r_i})
+\frac{GM_{\rm BH}\Delta m_i}{{}{\tilde r_i}}
-\nonumber \\ 
&\sum_{j>i}^N\!\left(
-\frac{GM_{r,j}\Delta m_j}{r_j}
+h_j\Delta m_j
+E_{\rm rot}^{(j)}\Delta m_j
\right),
\end{align}
where $h_j=\Gamma_j/(\Gamma_j-1)\,P_j/\rho_j${}{,} $E_{\rm rot}^{(j)}=\tfrac13\Omega_j^2 r_j^2${, and the summation extends over shells exterior to shell $i$}. {We take $\tilde r_i=r_{\rm LSO}$ for quasi-radial infall and $\tilde r_i=r_{\rm circ}$ when a disk forms.}

If $E_{\rm fb}^{(i)}>E_{\rm ub}^{(i)}$ (feedback with losses) or if the cumulative energy $E_{\rm fb}^{(<i)}$ exceeds $E_{\rm ub}^{(i)}$ (without losses), accretion halts. 
The model depends parametrically on the value of $\mathcal{M}_{\rm wind}$ and offers two possible modes to handle the feedback, with and without losses. The larger the value of $\mathcal{M}_{\rm wind}$, the stronger the BH feedback. Here we consider $\mathcal{M}_{\rm wind}\sim 3$ or $10$. For $\mathcal{M}_{\rm wind}\sim 3$ we reproduce the same efficiency parameter as in \citealt{Batta_2019arXiv190404835}, i.e., $\varepsilon\simeq 10^{-3}$, whereas $\mathcal{M}_{\rm wind}=10$ increases $\varepsilon$ to $\simeq 10^{-2}$. In the models with losses, a fraction of the BH energy released will be advected back onto it. Hence, this case, along with $\mathcal{M}_{\rm wind}=3$, estimates a lower bound of the ejected mass, $M_{\rm e}^{(1)}$ and an upper bound for the final BH mass ($M_{\rm BH}^{(1)}$) and spin ($a_{\rm BH}^{(1)}$) before the outer layers of the star are unbound. In contrast, models without losses and $\mathcal{M}_{\rm wind}=10$ provide an approximate lower bound for both the final $M_{\rm BH}^{(2)}$ and $a_{\rm BH}^{(2)}$, and a correspondingly larger estimate of the ejected mass $M_{\rm e}^{(2)}$. As illustrated in Fig.~\ref{fig:Max_final_a}, both $a_{\rm BH}^{(1)}$ and $a_{\rm BH}^{(2)}$ remain well below the dimensionless stellar spin because only part of the progenitor angular momentum can be incorporated into the BH during the collapse and subsequent accretion. The close overlap between $a_{\rm BH}^{(1)}(M_r)$ and $a_{\rm BH}^{(2)}(M_r)$ over most of the star indicates that both feedback prescriptions produce a very similar spin-up history while the same inner shells are being accreted. Noticeable differences appear only in the outermost layers, where the different feedback efficiencies mainly determine the mass coordinate at which the collapse is halted.

\begin{figure}
    \centering
    \includegraphics[width=\linewidth]{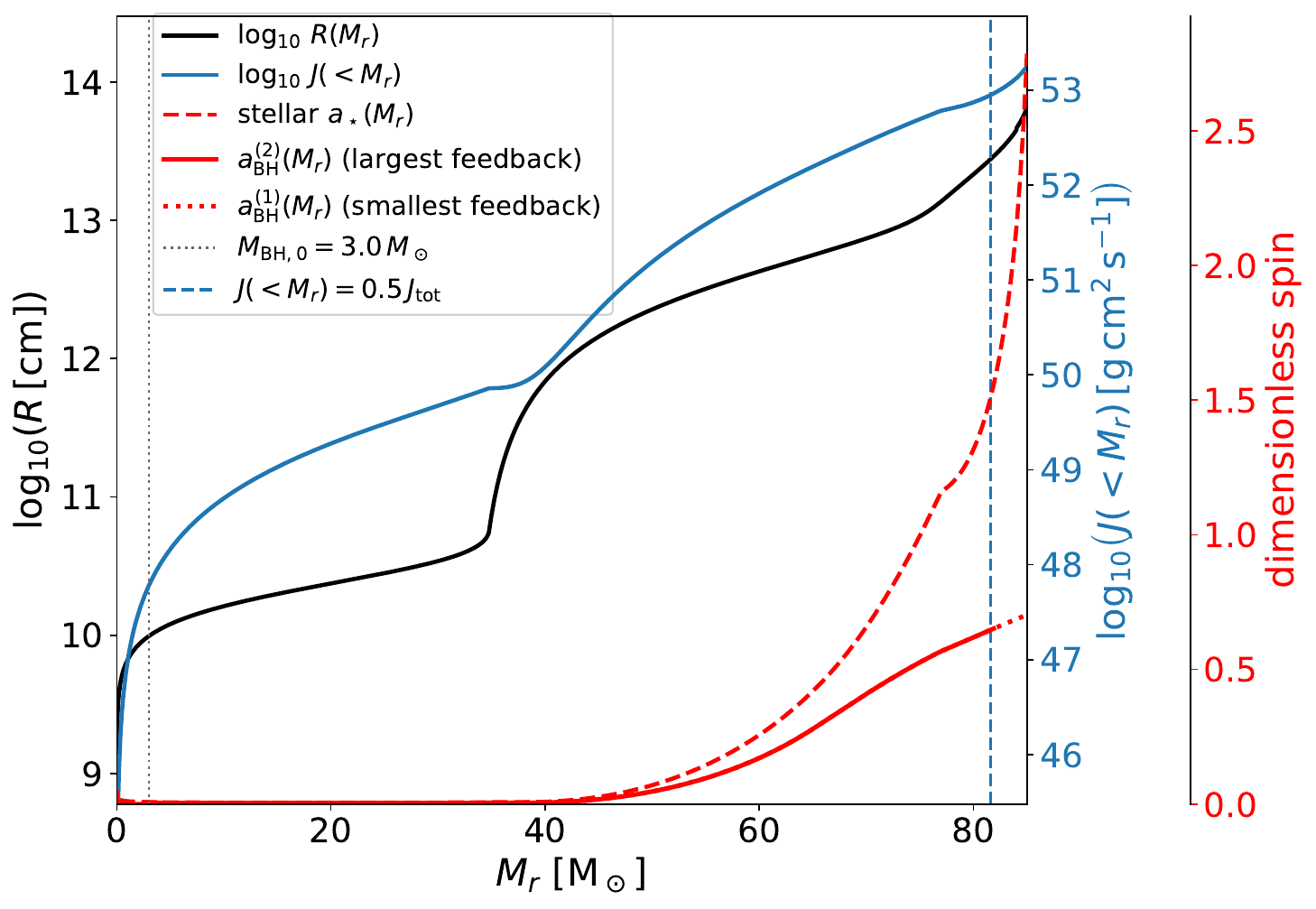}
    \caption{Radial structure and angular-momentum content of the stellar model with $M_{\rm ini}=85\,M_\odot$ and $\omega=0.15$ as functions of enclosed mass $M_r$, together with the corresponding stellar and black-hole spin parameters. The adopted initial BH mass, $M_{\rm BH,0}=3\,M_\odot$, is marked by the vertical dotted line. The black curve shows the radius $R(M_r)$, the blue curve the cumulative angular momentum $J(<M_r)$, the red dashed curve the stellar dimensionless spin $a_\star(M_r)$, and the red solid curve the BH spin $a_{\rm BH}^{(2)}(M_r)$ corresponding to the case with maximum feedback (no losses and $\mathcal{M}_{\rm wind}=10$), and the red dotted curve the BH spin $a_{\rm BH}^{(1)}(M_r)$ corresponding to the case with weaker feedback (with losses and $\mathcal{M}_{\rm wind}=3$). The vertical dashed blue line indicates the mass coordinate within which 50\% of the star's total angular momentum is enclosed.} 
    \label{fig:Max_final_a}
\end{figure}

\end{appendix}

\end{document}